\documentclass{article}
\usepackage{amsmath}
\usepackage{cite}
\usepackage{graphicx}
\usepackage{dcolumn}

\begin{document}

\date{}
\title{Equivalent potential-energy functions}
\author{Francisco M. Fern\'{a}ndez\thanks{%
fernande@quimica.unlp.edu.ar} \\
INIFTA, DQT, Sucursal 4, C. C. 16, \\
1900 La Plata, Argentina}
\maketitle

\begin{abstract}
Many researchers have proposed their own potential-energy functions for the
study of diatomic molecules, some of which have been proved to be
equivalent. In this paper we suggest the application of a few simple rules
to determine the equivalence between molecular potentials. The resulting
systematic procedure is illustrated by means of several examples.
\end{abstract}

\section{Introduction}

\label{sec:intro}

The starting point of most quantum-mechanical treatments of molecular
systems is the Born-Oppenheimer approximation\cite{BH54} that consists of
two steps. One first solves the Sch\"{o}dinger equation for the electrons
with the nuclei clamped at a set of conveniently chosen configurations. This
procedure yields the potential-energy surface. In the second step one solves
the Schr\"{o}dinger equation for the nuclei moving on the potential-energy
surface obtained in the first step. In the case of diatomic molecules one
obtains a potential-energy curve $V(r)$. For quite a long time,
spectroscopists have made use of suitable analytical expressions for $V(r)$%
\cite{H50} and now and again scientists propose other such functions.

Some of those analytical expressions for $V(r)$ have been proved to be
identical or equivalent. For example, Jia et al.\cite{JIAETAL12} proved the
equivalence between the Sun and Tietz\cite{T63} potentials and Liang et al.%
\cite{LTJ13} proved the equivalence between the Hua\cite{H90} and Tietz
potentials. There are so many potentials that it is not unlikely that other
such equivalences have passed unnoticed.

Sometimes it is not so easy to determine if two molecular potentials are
equivalent. The purpose of this paper is to propose a simple procedure for
comparing different potentials. The simplest way is to normalize them in the
same way.

\section{Normalization of molecular potentials}

\label{sec:normalization}

Suppose that $V(r)$ is the analytical expression for a molecular potential
that depends on some adjustable parameters. We propose to determine three of
those parameters by means of the conditions
\begin{equation}
V\left( r_{e}\right) =0,\;V^{\prime }\left( r_{e}\right)
=0,\;\lim\limits_{r\rightarrow \infty }V(r)=D_{e}.  \label{eq:V_conditions}
\end{equation}
Obviously, if $V^{\prime \prime }\left( r_{e}\right) >0$ then $r_{e}>0$ is
the equilibrium bond length and $D_{e}>0$ is the dissociation energy. The
resulting potential is positive definite: $V(r)\geq 0$ for all $0<r<\infty $.

We then write $V(r)$ as
\begin{equation}
V(r)=D_{e}\left[ 1-f(r)\right] ^{2},  \label{eq:V_f}
\end{equation}
from which we obtain two roots
\begin{equation}
f_{\pm }(r)=1\pm \sqrt{\frac{V(r)}{D_{e}}.}  \label{eq:f+-}
\end{equation}
Both yield the same potential but we choose $f_{-}(r)$ that behaves
asymptotically as $f(r\rightarrow \infty )=0$.

Let us begin with three simple cases. If we insert
\begin{equation}
V(r)=a+\frac{b}{r}+\frac{c}{r^{2}},  \label{eq:V_K_1}
\end{equation}
into equations (\ref{eq:V_conditions}) and solve for the adjustable
parameters we obtain
\begin{equation}
a=D_{e},\;b=-2D_{e}r_{e},\;c=D_{e}r_{e}^{2},  \label{eq:a,b,c_kratzer}
\end{equation}
that lead to the well-known Kratzer-Fues potential\cite{K20,F26}
\begin{equation}
V_{\text{K}}(r)=D_{e}\left( 1-\frac{r_{e}}{r}\right) ^{2}.  \label{eq:V_K_2}
\end{equation}

If we start with
\begin{equation}
V(r)=a+br^{-\alpha r}+ce^{-2\alpha r},\;\alpha >0,  \label{eq:V_M_1}
\end{equation}
we obtain
\begin{equation}
a=D_{e},\;b=-2D_{e}e^{\alpha r_{e}},\;c=D_{e}e^{2\alpha r_{e}},
\label{eq:a,b,c_Morse}
\end{equation}
and the Morse potential\cite{M29}
\begin{equation}
V_{\text{M}}(r)=D_{e}\left[ 1-e^{\alpha \left( r_{e}-r\right) }\right] ^{2}.
\label{eq:V_M_2}
\end{equation}
It follows from $V^{\prime \prime }\left( r_{e}\right) =k>0$ that
\begin{equation}
\alpha =\sqrt{\frac{k}{2D_{e}}}  \label{eq:alpha_M}
\end{equation}

The third example is
\begin{equation}
V(r)=a+\frac{b}{e^{\alpha r}+\beta }+\frac{c}{\left( e^{\alpha r}+\beta
\right) ^{2}},\;\alpha >0,  \label{eq:V_expo_1}
\end{equation}
where $\beta >-1$ in order to avoid an unphysical singularity in $r>0$. When
$\beta =0$ we obtain the Morse potential as a particular case. If we solve
equations (\ref{eq:V_conditions}) for $a$, $b$ and $c$ we obtain
\begin{equation}
a=D_{e},\;b=-2D_{e}\left( e^{\alpha r_{e}}+\beta \right) ,\;c=D_{e}\left(
e^{\alpha r_{e}}+\beta \right) ^{2},  \label{eq:a,b,c_expo}
\end{equation}
and
\begin{equation}
V(r)=D_{e}\left( 1-\frac{e^{\alpha r_{e}}+\beta }{e^{\alpha r}+\beta }%
\right) ^{2}.  \label{eq:V_expo_2}
\end{equation}
In this case we can obtain $\beta $ in terms of $\alpha $ from the condition
$V^{\prime \prime }\left( r_{e}\right) =k>0$
\begin{equation}
\beta =e^{\alpha r_{e}}\left( \alpha \sqrt{\frac{2D_{e}}{k}}-1\right) .
\label{eq:beta_expo}
\end{equation}
If we substitute this equation into $V^{\prime \prime \prime }\left(
r_{e}\right) =k_{3}$ we obtain an expression for $\alpha $
\begin{equation}
\alpha =\sqrt{\frac{2k}{D_{e}}}+\frac{k_{3}}{3k}.  \label{eq:alpha_expo}
\end{equation}
All the model parameters are completely determined by $D_{e}$,
$r_{e}$, $k$ and $k_{3}$ that are available from spectroscopic
data\cite{H50}.

In the next section we compare some analytical expressions for molecular
potentials proposed by some researchers in order to determine possible
equivalences.

\section{Examples of equivalent potentials}

\label{sec:Examples}

Two potential-energy functions are said to be equivalent if there exists a
bijective transformation of their parameters that maps one potential onto
the other, yielding the same function $V(r)$. Equivalence is therefore a
property of the function itself rather than of the particular
parametrization used to represent it. The normalization procedure merely
removes redundant parameters and recasts different potentials into a common
form suitable for direct comparison.

One of the most relevant potentials for present analysis was proposed by
Tietz\cite{T63} several years ago
\begin{equation}
V_{\text{T}}(r)=D_{e}+\frac{D_{e}\left[ \left( a+b\right) e^{-2\beta
r}-be^{-\beta r}\right] }{\left( 1+ce^{-\beta r}\right) ^{2}},\;\beta >0,
\label{eq:V_T_1}
\end{equation}
where $c\geq -1$ in order to avoid an unphysical singularity in $0<r<\infty $%
. This potential satisfies one of the normalization conditions $%
V(r\rightarrow \infty )=D_{e}$; from the remaining ones we obtain
\begin{equation}
a=e^{2\beta r_{e}}-2e^{\beta r_{e}}-c(c+2),\;b=2e^{\beta r_{e}}+2c,
\label{eq:a,b_Tietz}
\end{equation}
so that
\begin{equation}
V_{\text{T}}(r)=D_{e}\left( 1-\frac{e^{\beta r_{e}}+c}{e^{\beta r}+c}\right)
^{2}.  \label{eq:V_T_2}
\end{equation}
This expression is identical to our equation (\ref{eq:V_expo_2}) and to the
one derived by Jia et al.\cite{JIAETAL12}.

Hua\cite{H90} proposed the potential
\begin{equation}
U_{\text{H}}(r)=D_{e}\left[ \frac{1-e^{-b\left( r-r_{e}\right) }}{%
1-ce^{-b\left( r-r_{e}\right) }}\right] ^{2},\;b>0,  \label{eq:V_H_1}
\end{equation}
where $c\leq 1$. This expression satisfies all the normalization conditions (%
\ref{eq:V_conditions}) so that it only remains to obtain the function $f(r)$%
:
\begin{equation}
f(r)=\frac{e^{br_{e}}\left( 1-c\right) }{e^{br}-ce^{br_{e}}}.
\label{eq:f(r)_Hua}
\end{equation}
If we define $c^{\prime }=-ce^{br_{e}}$, we obtain
\begin{equation}
U_{\text{H}}(r)=D_{e}\left( 1-\frac{e^{br_{e}}+c^{\prime }}{\mathbf{e}%
^{br}+c^{\prime }}\right) ^{2},\;c^{\prime }\geq -1,  \label{eq:V_H_2}
\end{equation}
that is equivalent to the Tietz potential. As stated in the introduction of
present paper Jia et al.\cite{JIAETAL12} already proved this equivalence.

Badawi et al\cite{BBB72} proposed a general potential of the form
\begin{equation}
V_{\text{BBB}}{\left( r\right) }=M\left( \frac{\exp {\left( -\gamma r\right)
}}{1+a\exp {\left( -\gamma r\right) }}\right) ^{2}+\frac{N\exp {\left(
-\gamma r\right) }}{1+a\exp {\left( -\gamma r\right) }}+L.  \label{eq:V_BBB}
\end{equation}
It follows from equation (\ref{eq:V_conditions}) that
\begin{equation}
M=D_{e}\left( e^{\gamma r_{e}}+a\right) ^{2},\;N=-2D_{e}\left( e^{\gamma
r_{e}}+a\right) ,\;L=D_{e},  \label{eq:M,N,L}
\end{equation}
and the potential becomes
\begin{equation}
V_{\text{BBB}}(r)=D_{e}\left( 1-\frac{e^{\gamma r_{e}}+a}{e^{\gamma r}+a}%
\right) ^{2},  \label{eq:V_BBB_2}
\end{equation}
that is identical to the Tietz potential.

\c{S}im\c{s}ek and \"{O}z\c{c}elik\cite{SO94} proposed the four-parameter
exponential type potential(FPEP)
\begin{equation}
V_{\text{FPEP}}(r)=\frac{Ae^{\lambda r}}{C+e^{\lambda r}}-\frac{Be^{\lambda
r}}{\left( C+e^{\lambda r}\right) ^{2}},\;A,B>0.  \label{eq:V_FPEP}
\end{equation}
In order to make this potential positive definite we add a constant $V_{0}$
that does not change the physical features of the model and consider $U_{%
\text{FPEP}}(r)=V_{0}+V_{\text{FPEP}}(r)$. It follows from the normalization
conditions (\ref{eq:V_conditions}) that

\begin{equation}
V_{0}=\frac{D_{e}e^{2\lambda r_{e}}}{C^{2}},\;A=\frac{D_{e}\left(
C^{2}-e^{2\lambda r_{e}}\right) }{C^{2}},\;B=\frac{D_{e}\left( e^{\lambda
r_{e}}+C\right) ^{2}}{C},  \label{eq:V0,A,B_FPEP}
\end{equation}
and
\begin{equation}
U_{\text{FPEP}}(r)=D_{e}\left( 1-\frac{e^{\lambda r_{e}}+C}{e^{\lambda r}+C}%
\right) ^{2},  \label{eq:V_FPEP_2}
\end{equation}
that is the Tietz potential.

Jia et al.\cite{JIAETAL01} introduced a deformed five-parameter
exponential-type potential (DFPEP)
\begin{equation}
V_{\text{DFPEP}}(r)=P_{1}+\frac{P_{2}}{e^{2\alpha r}+q}+\frac{P_{3}}{\left(
e^{2\alpha r}+q\right) ^{2}},\;\alpha >0,  \label{eq:V_DFPEP_1}
\end{equation}
where $q\geq -1$. From the conditions (\ref{eq:V_conditions}) we obtain
\begin{equation}
P_{1}=D_{e},\;P_{2}=-2D_{e}\left( e^{2\alpha r_{e}}+q\right)
,\;P_{3}=D_{e}\left( e^{2\alpha r_{e}}+q\right) ^{2},
\label{eq:P1,P2,P3_DFPEP}
\end{equation}
and
\begin{equation}
V_{\text{DFPEP}}(r)=D_{e}\left( 1-\frac{e^{\beta r_{e}}+q}{e^{\beta r}+q}%
\right) ^{2},\;\beta =2\alpha ,  \label{eq:V_DFPEP_2}
\end{equation}
that is equivalent to the normalized Tietz potential (\ref{eq:V_T_2}).

Arda and Sever\cite{AS11} proposed the generalized Morse (GM) potential
\begin{equation}
V_{\text{GM}}(r)=V_{1}e^{-2\beta \frac{r-r_{0}}{r_{0}}}-V_{2}e^{-\beta \frac{%
r-r_{0}}{r_{0}}}.  \label{eq:V_GM}
\end{equation}
From $V^{\prime }\left( r_{e}\right) =0$ and $V\left( r_{e}\right) =-D_{e}$
we obtain
\begin{equation}
V_{1}=D_{e}e^{2\beta \left( \frac{r_{e}}{r_{0}}-1\right)
},\;V_{2}=2D_{e}e^{\beta \left( \frac{r_{e}}{r_{0}}-1\right) },
\label{eq:V1,V2,GM}
\end{equation}
and
\begin{equation}
V_{\text{GM}}(r)=D_{e}\left[ e^{2\alpha \left( r_{e}-r\right) }-2e^{\alpha
\left( r_{e}-r\right) }\right] ,\;\alpha =\frac{\beta }{r_{0}},
\label{eq:V_GM_2}
\end{equation}
that is the standard form of the Morse potential. We can rewrite $V_{\text{GM%
}}(r)+D_{e}$ in the form (\ref{eq:V_M_2}).

Falaye et al.\cite{FIH15} proposed the shifted Tietz-Wei (sTW) oscillator
\begin{equation}
V_{\text{sTW}}(r)=V_{e}\frac{2(c_{h}-1)e^{-b_{h}\left( r-r_{e}\right)
}-\left( c_{h}^{2}-1\right) e^{-2b_{h}\left( r-r_{e}\right) }}{\left[
1-c_{h}e^{-b_{h}\left( r-r_{e}\right) }\right] ^{2}},\;b_{h}>0.
\label{eq:V_sTW_1}
\end{equation}
This potential satisfies $V\left( r_{e}\right) =-V_{e}$, $V^{\prime }\left(
r_{e}\right) =0$ and $V(r\rightarrow \infty )=0$; therefore $U(r)=V(r)+V_{e}$
is positive definite. It follows from equation (\ref{eq:f+-}) that
\begin{equation}
f(r)=\frac{e^{b_{h}r_{e}}\left( 1-c_{h}\right) }{%
e^{b_{h}r}-c_{h}e^{b_{h}r_{e}}}=\frac{e^{b_{h}r_{e}}+c}{e^{b_{h}r}+c}%
,\;c=-c_{h}e^{b_{h}r_{e}},  \label{eq:f_sTW}
\end{equation}
and
\begin{equation}
U_{\text{sTW}}(r)=D_{e}\left( 1-\frac{e^{b_{h}r_{e}}+c}{e^{b_{h}r}+c}\right)
^{2},\;D_{e}=V_{e}.  \label{eq:V_sTW_2}
\end{equation}
Clearly, the sTW oscillator is equivalent to the Tietz potential.

Onate et al.\cite{OOOI20} proposed the q-deformed six-parameter
exponential-type potential model
\begin{equation}
V_{\text{qDSPEP}}(r)=Q_{1}-\frac{Q_{2}}{1-qe^{-2\alpha r}}+\frac{%
Q_{3}e^{-2\alpha r}}{1-qe^{-2\alpha r}}-\frac{Q_{4}e^{-4\alpha r}}{\left(
1-qe^{-2\alpha r}\right) ^{2}},\;\alpha >0.  \label{eq:V_six-param}
\end{equation}
If we solve the normalization conditions (\ref{eq:V_conditions}) for $Q_{2}$%
, $Q_{3}$ and $Q_{4}$ we obtain
\begin{equation}
Q_{2}=Q_{1}-D_{e},\;Q_{3}=q\left( Q_{1}+D_{e}\right) -2D_{e}e^{2\alpha
r_{e}},\;Q_{4}=-D_{e}\left( e^{2\alpha r_{e}}-q\right) ^{2},
\label{eq:Q_i_six-param}
\end{equation}
and
\begin{equation}
V_{\text{qDSPEP}}(r)=D_{e}\left( 1-\frac{e^{\beta r_{e}}+c}{e^{\beta r}+c}%
\right) ^{2},\;\beta =2\alpha ,\;c=-q,  \label{eq:V_six-param_2}
\end{equation}
that is the Tietz potential.

Okorie et al.\cite{OIC20} proposed the deformed exponential-type potential
(DEP)
\begin{equation}
U_{\text{DEP}}(r)=P_{1}+\frac{P_{2}e^{-2\alpha \left( r-r_{0}\right) }}{%
1-qe^{-2\alpha \left( r-r_{0}\right) }}+\frac{P_{3}e^{-4\alpha \left(
r-r_{0}\right) }}{\left[ 1-qe^{-2\alpha \left( r-r_{0}\right) }\right] ^{2}}.
\label{eq:U_DEP}
\end{equation}
If we define $P_{2}^{\prime }=P_{2}e^{2\alpha r_{0}}$,$\;P_{3}^{\prime
}=P_{3}e^{4\alpha r_{0}}$ and $q\prime =qe^{2\alpha r_{0}}$ then we realize
that this potential is equivalent to the DFPEP (\ref{eq:V_DFPEP_1}) and that
$r_{0}$ is an irrelevant parameter with no effect on the physics of the
problem. In spite of this fact, Okorie et al. managed to obtain an
expression for $r_{0}$. Although the normalization of the potential (\ref
{eq:U_DEP}) follows directly from the results obtained previously for (\ref
{eq:V_DFPEP_1}), it may be illustrative to carry out the normalization of
the model (\ref{eq:U_DEP}). It follows from the conditions (\ref
{eq:V_conditions}) that
\begin{equation}
P_{1}=D_{e},\;P_{2}=2D_{e}e^{-2\alpha r_{0}}\left( qe^{2\alpha
r_{0}}-e^{2\alpha r_{e}}\right) ,P_{3}=D_{e}e^{-4\alpha r_{0}}\left(
qe^{2\alpha r_{0}}-e^{2\alpha r_{e}}\right) ^{2}\;,  \label{eq:Pi_DEP}
\end{equation}
and
\begin{equation}
U_{\mathrm{DEP}}(r)=D_{e}\left( 1-\frac{e^{\beta r_{e}}+c}{e^{\beta r}+c}%
\right) ^{2},\;c=-qe^{2\alpha r_{0}},\;\beta =2\alpha \text{.}
\label{eq:U_DEP_2}
\end{equation}
Clearly, $U_{\text{DEP}}(r)$ is another potential that is equivalent to the
Tietz one. Although Eyube et al.\cite{ETY22} corrected some flaws in the
paper by Okorie et al.\cite{OIC20}, they still retained the utterly
ineffective parameter $r_{0}$ in their equations.

\section{Asymptotic behaviour}

\label{sec:Asymptotic}

Finally, a few words on the asymptotic behaviour of the molecular potential.
It is well known that the molecular potential $V(r)$ should tend to plus
infinity at origin as $r^{-1}$ because of the Coulomb repulsion of the nuclei%
\cite{H50}. On the other hand, $D_{e}-V(r)$ should tend to zero as $r^{-6}$%
\cite{H50}. Although these exponential potentials are useful
phenomenological models in the vicinity of the equilibrium distance, they do
not reproduce the correct asymptotic behaviour at either short or long
internuclear distances. It has been shown that a rational function is a
straightforward way to obtain molecular potentials with suitable asymptotic
behaviour\cite{OT18, TO22, OT23}.

\section{Conclusions}

\label{sec:conclusions}

In this paper we have proposed a simple and systematic normalization
procedure for the comparison of analytical potential-energy functions used
in the study of diatomic molecules. The method consists of determining three
adjustable parameters from the conditions (\ref{eq:V_conditions}), and then
rewriting the resulting potential in the form (\ref{eq:V_f}). This
representation greatly facilitates the direct comparison between different
models.

The normalization procedure shows that several molecular potentials proposed
in the literature are not genuinely different models but merely alternative
parametrizations of the same function $V(r)$. In particular, the Hua, DFPEP,
shifted Tietz-Wei, qDSPEP, and DEP potentials can all be transformed into
the same normalized form of the Tietz potential by means of suitable
bijective parameter transformations. This analysis also reveals the
existence of redundant parameters, such as the parameter $r_{0}$ appearing
in the GM and DEP models.

The examples discussed throughout the paper illustrate that potential
equivalence is a property of the functional form of $V(r)$ itself rather
than of the particular parametrization chosen to represent it. In this
context, normalization does not alter the physics of the problem but simply
removes redundant degrees of freedom and recasts different models into a
common representation. We have also shown that the normalized function $f(r)$
provides a convenient framework for relating the parameters of the model
potential to spectroscopic quantities such as the force constant and the
anharmonicity coefficient. At the same time, the analysis of the asymptotic
behaviour indicates that these exponential potentials do not reproduce the
correct short- and long-range limits expected for realistic molecular
interactions, although they may still provide useful phenomenological
descriptions near the equilibrium internuclear distance.

The present approach may help identify unnoticed equivalences between
apparently different molecular potentials, simplify their classification,
and avoid the introduction of ostensibly new models that are functionally
identical to previously known potentials.


\begin{thebibliography}{99}
\bibitem{BH54}  M. Born and K. Huang, Dynamical Theory of Crystal Lattices
(Oxford University Press, New York, 1954).

\bibitem{H50}  G. Herzberg, Molecular Spectra and Molecular Structure. I.
Spectra of Diatomic Molecules, Second ed. (Van Nostrand Reinhold, New York,
1950).

\bibitem{JIAETAL12}  C-S. Jia, Y-F. Diao, X-J. Liu, P-Q. Wang, J-Y. Liu, and
G-D. Zhang, J. Chem. Phys. \textbf{137}, 014101 (2012).

\bibitem{T63}  T. Tietz, J. Chem. Phys. \textbf{38}, 3036 (1963).

\bibitem{LTJ13}  G.-C. Liang, H.-M. Tang, and C.-S. Jia, Comp: Theor. Chem.
\textbf{1020}, 170 (2013).

\bibitem{H90}  W. Hua, Phys. Rev. A \textbf{42}, 2524 (1990).

\bibitem{K20}  A. Kratzer, Z. Physik \textbf{3}, 289 (1920).

\bibitem{F26}  E. Fues, Ann. Phys. \textbf{386}, 281 (1926).

\bibitem{BBB72}  M. Badawi, N. Bessis, and G. Bessis, J. Phys. B \textbf{5},
L157 (1972).

\bibitem{M29}  P. M. Morse, Phys. Rev. \textbf{34}, 57 (1929).

\bibitem{SO94}  M. \c{S}im\c{s}ek and S. \"{O}z\c{c}elik, Phys. Lett. A
\textbf{186}, 35 (1994).

\bibitem{JIAETAL01}  C.-S. Jia, Y. Zhang, X.-L. Zeng, and L.-T. Sun, Commun.
Theor. Phys. (Beijing, China) \textbf{36}, 641 (2001).

\bibitem{AS11}  A. Arda and R. Sever, Commun. Theor. Phys. \textbf{56}, 51
(2011).

\bibitem{FIH15}  B. J. Falaye, S. M. Ikhdair, and M. Hamzavi, J. Theor.
Appl. Phys. \textbf{9}, 151 (2015).

\bibitem{OOOI20}  C. A. Onate, M. C. Onyeaju, U. S. Okorie, and A. N. Ikot,
Res. Phys. \textbf{16}, 102959 (2020).

\bibitem{OIC20}  U. S. Okorie, A. N. Ikot, and E. O. Chukwuocha, Bull.
Korean Chem. Soc. \textbf{41}, 609 (2020).

\bibitem{ETY22}  E. S. Eyube, P. Timtere, and J. B. Yerima, Can. J. Phys.
\textbf{100}, 351 (2022).

\bibitem{OT18}  H. O. Olivares-Pil\'{o}n and A. V. Turbiner, Ann. Phys.
\textbf{393}, 335 (2018).

\bibitem{TO22}  A. V. Turbiner and H. O. Olivares-Pil\'{o}n, Mol. Phys.
\textbf{120}, e2064784 (2022).

\bibitem{OT23}  H. O. Olivares-Pil\'{o}n and A. V. Turbiner, J. Phys. B
\textbf{56}, 165101 (2023).
\end{thebibliography}
\end{document}